\documentclass{aa}  
\usepackage{graphicx}
\usepackage{natbib}
\usepackage{derivative}
\usepackage{xcolor}
\usepackage{float}
\usepackage{rotating}
\usepackage{booktabs}
\usepackage{adjustbox}
\usepackage{multicol}
\usepackage{multirow}
\usepackage{makecell}

\usepackage{txfonts}

\newcommand{\PACF}{P_\mathrm{ACF}}
\newcommand{\HACF}{H_\mathrm{ACF}}
\newcommand{\GACF}{G_\mathrm{ACF}}
\newcommand{\PGWPS}{P_\mathrm{GWPS}}
\newcommand{\PCS}{P_\mathrm{CS}}
\newcommand{\HCS}{H_\mathrm{CS}}
\newcommand{\prot}{$P_\text{rot}$}
\newcommand{\sph}{$S_\text{\!ph}$}
\newcommand{\kep}{\textit{Kepler}}

\begin{document}

   \title{ROOSTER: a machine-learning analysis tool for \textit{Kepler} stellar rotation periods}

   \author{S.N.~Breton\inst{1}
          \and
          A.R.G.~Santos\inst{2}
          \and
          L.~Bugnet\inst{1}
         \and
          S.~Mathur\inst{3,4}
          \and
          R.A~García\inst{1}
          \and
          P.L.~Pallé\inst{3,4}
          }

   \institute{AIM, CEA, CNRS, Université Paris-Saclay, Université Paris Diderot, Sorbonne Paris Cité, F-91191 Gif-sur-Yvette, France\\  \email{sylvain.breton@cea.fr}
         \and
             Space Science Institute, 4765 Walnut Street, Suite B, Boulder CO 80301, USA
        \and
        Instituto de Astrof\'{i}sica de Canarias, 38200 La Laguna, Tenerife, Spain
        \and
        Universidad de La Laguna, Dpto. de Astrof\'{i}sica, 38205 La Laguna, Tenerife, Spain
            }

   \date{}

 \abstract{
   In order to understand stellar evolution, it is crucial to efficiently determine stellar surface rotation periods. Indeed, while they are of greatest importance in stellar models, angular momentum transport processes inside stars are still poorly understood today. Surface rotation, linked to the age of the star, is one of the constraints needed to improve the way those processes are modelled. Statistics of the surface rotation periods for a large sample of stars of different spectral types are thus necessary.
   An efficient tool to automatically determine reliable rotation periods is needed when dealing with large samples of stellar photometric datasets. The objective of this work is to develop such a tool.
   For this purpose, machine learning classifiers represent relevant bricks for the basis of our new methodology. Random forest learning abilities are exploited to automate the extraction of rotation periods in \textit{Kepler} light curves. Rotation periods and complementary parameters are obtained from three different methods: a wavelet analysis, the autocorrelation function of the light curve, and the composite spectrum. We train three different classifiers: one to detect if rotational modulations are present in the light curve, one to flag close binary or classical pulsators candidates that can bias our rotation period determination, and finally one classifier to provide the final rotation period.
   We test our machine learning pipeline on 23,431 stars of the \textit{Kepler} K and M dwarf reference rotation catalog of \citet{2019ApJS..244...21S} for which 60\% of the stars have been visually inspected. For the sample of 21,707 stars where all the input parameters are provided to the algorithm, 94.2\% of them are correctly classified (as rotating or not). Among the stars that have a rotation period in the reference catalog, the machine learning provides a period that agrees within 10\% of the reference value for 95.3\% of the stars.
   Moreover, the yield of correct rotation periods is raised to 99.5\% after visually inspecting 25.2\% of the stars. Over the two main analysis steps, rotation classification and period selection, the pipeline yields a global agreement with the reference values of 92.1\% and 96.9\% before and after visual inspection.
   Random forest classifiers are efficient tools to determine reliable rotation periods in large samples of stars. The methodology presented here could be easily adapted to extract surface rotation periods for stars with different spectral types or observed by other instruments such as K2, TESS or, in a near future, PLATO.}
   
   \maketitle
%

\section{Introduction}

Low-mass stars with convective outer layers (hereafter solar-type stars) can exhibit magnetic activity \citep[e.g.][]{Brun2017}. In the Sun, one of the manifestations of magnetic activity is the emergence of magnetic spots at its surface. While for stars other than the Sun one cannot directly image starspots in great detail, one can in principle detect their impact on, for example, stellar brightness. As the star rotates, starspots come in and out of the visible hemisphere, thus modulating the stellar brightness. As a consequence, such spot modulation encloses information on stellar magnetic activity and surface rotation \citep[e.g.][]{2005LRSP....2....8B,2009A&ARv..17..251S}.

Surface rotation periods represent a key parameter to understand stellar angular momentum transport. This process is not yet sufficiently understood to be correctly implemented in stellar evolution codes \citep[e.g.][and references therein]{2019ARA&A..57...35A}. Neglecting angular momentum transport may have significant impact on the stellar age estimates \citep[e.g.][]{2010A&A...509A..72E}. These stellar ages are crucial for studying the evolution of the Milky Way \citep[e.g.][]{2013MNRAS.429..423M}, as well as for a better characterization of planetary systems \citep[e.g.][]{2018ASSP...49..119H}. The evolution of planetary systems, driven by tidal and magnetic effects, is indeed strongly influenced by the stellar rotation rate, through angular momentum exchange between planets and their host star star \citep[e.g.][]{2014ApJ...787..131Z,2015A&A...580L...3M,2016CeMDA.126..275B,2017ApJ...847L..16S,2019A&A...621A.124B}.

The surface rotation period is found to be a strong function of the stellar age: low-mass stars with an external convective envelope spin-down during their main-sequence evolution. The first empirical relation between the two stellar parameters was proposed by \citet[][also known as Skumanich spin-down law]{1972ApJ...171..565S}. In particular, for young main-sequence solar-type stars, the rotation period can be used to constrain stellar ages through the so-called gyrochronology relations \citep[e.g.][]{Barnes2003,Barnes2007,2008ApJ...687.1264M,2011ApJ...733L...9M,2015Natur.517..589M,2014A&A...572A..34G}. However, for stars older than the Sun, the ages determined by gyrochronology deviate from the asteroseismic ages \citep[][]{2015MNRAS.450.1787A,2016Natur.529..181V}. \citet{2020arXiv200509387A} also showed that the empirical gyrochronology relations cannot reproduce the stellar ages inferred from velocity dispersion. These discrepancies justify the need for improved gyrochronology relations. Furthermore, the presence of surface rotational modulation in the light curves has a limiting effect on the detection of low-amplitude transiting exoplanets \citep[e.g.][]{Cameron2017}. The amplitude of those rotational modulations is directly linked to the surface stellar magnetic activity level and leads to difficulties to perform asteroseismic analysis of active solar-like stars because this magnetism inhibits pulsations \citep[e.g.][and references therein]{2019LRSP...16....4G,2019FrASS...6...46M}. 

During its four-year nominal mission, the \kep\ satellite \citep{2010Sci...327..977B} collected high-quality, long-term, and nearly continuous photometric data for almost 200,000 targets \citep{2017ApJS..229...30M} in the Cygnus-Lyra region. After the failure of two reaction wheels, \kep\ was reborn as a new mission called K2 \citep{2014PASP..126..398H}. K2 observed more than 300,000 targets around the ecliptic \citep{2016ApJS..224....2H} over 20 three-month campaigns. The Transiting Exoplanet Survey Satellite \citep[TESS,][]{2015JATIS...1a4003R} observed nearly the all sky during its nominal mission. TESS gathered data for tens of millions of stars with an observational length ranging from 27 days to one year. This large amount of observations represent the best photometric dataset to study the distribution of stellar surface rotation periods, $P_\text{rot}$. But to reach this goal and extract periods for large samples of stars, automatic procedures are required. Such an attempt was recently undertaken with a deep learning approach using convolutional neural networks \citep{2020arXiv200509682B}. However, training those networks requires a heavy computational power. Hence, the development of an easy-to-train machine learning procedure coupled with the outputs of widely-used rotation pipelines will be an extremely valuable asset. Indeed, this is the objective of the current study.

Random forest (RF) algorithms \citep[][]{Breiman2001} are either able to classify large samples of stars (when used for classification) or to provide parameter estimates (when used for regression). In asteroseismology, RF algorithms have already been used to estimate stellar surface gravities \citep[$\log g$,][]{2018A&A...620A..38B} and to automatically recognise solar-like pulsators  \citep{2019A&A...624A..79B}.
They have also recently been used to perform analyses linked to surface stellar rotation, but only to infer long rotation periods from TESS 27-day-long light curves \citep{2020AJ....160..168L}.
In this work, we present the \textit{Random fOrest Over STEllar Rotation} (ROOSTER), which is designed to select a rotation period for stars observed by \textit{Kepler} through a combination of RF classifiers applied to a variety of methods used to extract $P_\text{rot}$ and different ways to correct the light curves. 
Thus, the main goal of ROOSTER is to automatically achieve a large degree of reliability that could even be improved by performing a small number of human visual validation of certain results, which are also suggested by the pipeline.

The layout of the paper is as follows. Section~\ref{section:data_sample} presents the sample of K and M stars used in this work as well as the methodology to extract the surface rotation periods. Section~\ref{section:parameters} describes all the parameters used to feed ROOSTER.
In Sect.~\ref{section:pipeline}, we explain how the ROOSTER pipeline is built and we detail the training scheme we chose for this work. In Sect.~\ref{sec:results}, we present the rotation periods obtained with the ROOSTER pipeline and the strategy implemented for stars with missing input parameters. Results, limitations, and possibilities to increase the accuracy of ROOSTER are discussed in Sect.~\ref{section:discussion}. Finally, conclusions and perspectives are given in Sect.~\ref{section:conclusion}.


\section{Observations and methodology to extract surface rotation periods \label{section:data_sample}}

The target sample of this work is comprised of main-sequence K and M stars observed by the \kep\ mission. This sample was the one analysed in the rotation study done by \citet{2019ApJS..244...21S}. It is the most complete rotation catalog for the \kep\ K and M dwarfs. Hereafter, this catalog will be referred as S19. In it, 26,521 targets were selected according to the \kep\ Stellar Properties Catalog for Data Release 25 \citep[DR25][]{2017ApJS..229...30M} and for 15,290 of them, a $P_\mathrm{rot}$ was provided. As the purpose of this work is rotation analysis, different types of potential contaminants are removed from our study, namely \textit{Kepler} objects of interest, eclipsing binaries, misclassified red giants, misclassified RR Lyrae, light curves with photometric pollution by nearby stars or multiple signals, classical pulsators (namely $\delta$ Scuti and/or $\gamma$ Doradus) candidates, and contact binary candidates (see S19 for details on the contaminants).   

\noindent S19 determined average rotation periods, \prot, and average photometric activity indexes \citep[\sph,][]{2014A&A...562A.124M}, for more than $60\%$ of the targets. In fact, about $30\%$ of the reported \prot\ by S19 are new detections in comparison with the previous largest \prot\ catalog \citep{2014ApJS..211...24M}. In addition to an automatic selection of reliable \prot, the S19 analysis required an extensive amount of visual examinations (see Section~\ref{technique}), which justifies the need for a pipeline such as ROOSTER.



\subsection{Light curve preparation \label{subsection:data_calibration}}

We use three sets of KEPSEISMIC light curves\footnote{KEPSEISMIC data are available at MAST via \url{http://dx.doi.org/10.17909/t9-mrpw-gc07}.}, the same used in S19. KEPSEISMIC data is calibrated using KADACS \citep[\kep\ Asteroseimic Data Analysis and Calibration Software;][]{2011MNRAS.414L...6G}, which applies customized apertures and corrects for outliers, jumps, drifts, and discontinuities at the edges of the \kep\ Quarters. 
Because regular interruptions in the observations (such as the gaps induced every three days due to the angular momentum dump of the \textit{Kepler} satellite) could produce a false detection of $P_\text{rot}$, gaps
shorter than 20 days are filled by implementing inpainting techniques using a multi-scale discrete cosine transform \citep{2014A&A...568A..10G,2015A&A...574A..18P}.

We use in particular three different data sets obtained with three high-pass filters, with cutoff periods at 20, 55, and 80 days. The transfer function is unity for periods shorter than the cutoff period, while for longer periods it varies sinusoidaly, slowly approaching zero at twice the cutoff period. Thus, for a given filter, it is possible to recover periods longer than the respective cutoff period. The main goal of using and comparing three different filters is to find the best trade-off between filtering the undesired instrumental modulations at long periods while keeping the intrinsic stellar rotational signal. KEPSEISMIC light curves are optimized for asteroseismic studies, but are also very well suited for rotation and magnetic activity studies (see for example appendix B in S19).

\subsection{Methodology to extract $P_\text{rot}$ \label{technique}}

Surface rotation periods can be studied with different methods, namely: periodogram analysis \citep[e.g.][]{Reinhold2013a,2013A&A...557L..10N}; Gaussian processes \citep[e.g.][]{2018MNRAS.474.2094A}; gradient power spectrum analysis \citep[e.g.][]{2020A&A...633A..32S,2020A&A...636A..69A}; autocorrelation function \citep[ACF, e.g.][]{2013ApJ...775L..11M,2014ApJS..211...24M}; time-period analysis based on wavelets \citep[e.g.][]{2010A&A...511A..46M,2014A&A...572A..34G}; or with a combination of different diagnostics such as the composite spectrum \citep[CS, e.g.][]{2016MNRAS.456..119C,2017A&A...605A.111C,2019ApJS..244...21S}. Using artificial data, \citet{2015MNRAS.450.3211A} compared the performance of different pipelines used to determine the rotation. 
The authors found that the combination of the data set preparation and different rotation diagnostics, in particular the one implemented in this work (see next paragraph), yields the best set of results in terms of reliability and completeness. 


In this work, we use the procedure followed in S19 to extract $P_\text{rot}$ combining three analysis methods (time-period analysis, ACF, and CS) with the KEPSEISMIC light curves. 
Those $P_\text{rot}$ are respectively denoted $\PACF$, $\PGWPS$ and $\PCS$. Control parameters related to the ACF and CS are also extracted, they are denoted $\HACF$, $\GACF$, and $\HCS$. Each method also allows us to compute a magnetic activity proxy \sph. Description and details concerning those parameters can be found in Appendix~\ref{appendix:our_methodology}. 

During the visual examination in S19, a group of classical pulsators (CP) and close binaries (CB) candidates was identified. Those stars are referred hereafter as Type~1 CP/CB candidates. These targets are typically fast rotators ($P_\text{rot}<7$ days), showing very stable and fast beating patterns, a large number of harmonics in the power spectrum, and/or very large \sph\ values. Interestingly, \citet{Simonian2019} found that the majority of the fast rotators observed by \kep\ are tidally-synchronized binaries. In fact, there is a significant overlap between the Type~1 CP/CB candidates and the tidally-synchronized binaries (see S19). This suggests that the modulation seen in the light curve of such objects may still be related to rotation but not of single targets. Therefore, we advice caution while studying such targets. In S19, the Type~1 CP/CB candidates were flagged through visual inspection. 


\section{Parameters used to feed the machine learning algorithms \label{section:parameters}}

To ensure that ROOSTER is able to identify targets with detectable rotation modulation and to extract reliable rotation periods, the selection of relevant parameters is crucial. They will help the algorithm to perform successive classifications as described in Sect.~\ref{section:pipeline}. Indeed, a random forest algorithm relies on the following principle: during the training step, a forest of classification trees is grown to perform splits between the elements with different labels \citep{Breiman2001}. From the root of the tree, the training sample is progressively separated in purer (considering the labels) sub-samples. This task is performed by selecting at each iteration the best split over a range of randomly generated range of splits using the input parameters. 

The three rotation periods ($\PACF$, $\PGWPS$, and $\PCS$) obtained thanks to the method described in Sect.~\ref{technique} and in Appendix~\ref{appendix:our_methodology}, combined with the three sets of data corresponding to the three different high-pass filtered KEPSEISMIC light curves (20, 55 and 80 days), yield a total of 9 rotation-period candidates for each star. The objective is to train the algorithm to choose the rotation period given in S19. To do so, we also feed our pipeline with the parameters (amplitude, central value and standard deviation) of the five first Gaussian functions fitted on the GWPS and the CS. If, for a given star, the $i^\mathrm{th}$ Gaussian function could not be fitted, the corresponding parameters are set to zero. We also include the $\chi^2$ of the two fits, the number of Gaussian functions fitted to each GWPS and CS and the mean level of noise. This makes 108 additional input parameters (36 for each filter). As only a few stars have a $6^\mathrm{th}$ Gaussian function fitted for both GWPS and CS method, we verified that keeping the corresponding parameters out of the classification does not affect the result.   

We also feed the random forest algorithm with the $\HACF$, $\GACF$, and $\HCS$ for each filter, and the \sph\ with corresponding errors for each method and each filter, leading to 27 more parameters.

Because the rotation periods depend on the global parameters of the stars (age, spectral type, etc), we complement our set of input parameters with the effective temperature, $T_\mathrm{eff}$, and logarithm of surface gravity $\log g$ from the DR25 catalog \citep{2017ApJS..229...30M}.
We also decide to use the Flicker in Power metric \citep[FliPer, ][]{2018A&A...620A..38B}, which is correlated with the stellar surface gravity. FliPer is a measure of the total power in the stellar power spectral density (PSD) between a low frequency cut-off, $\nu_\mathrm{C}$ and the Nyquist frequency, $\nu_\mathrm{N}$. It can be defined as:  

\begin{equation}
    \mathrm{FliPer} \; (\nu_\mathrm{C}) = \frac{1}{\nu_\mathrm{N} - \nu_\mathrm{C}} \int_{\nu_\mathrm{C}}^{\nu_\mathrm{N}} \mathrm{PSD} (\nu) \; \mathrm{d}\nu - P_n \;,
\end{equation}
where $P_n$ is an estimate of the photon noise level which depends on the magnitude of the star. In this work, we used a noise estimation calibrated with the whole set of KEPSEISMIC calibrated light curves. We consider the following four FliPer values: $F_{0.7}$, $F_{7}$, $F_{20}$, $F_{50}$ for $\nu_\mathrm{C}$ of 0.7, 7, 20 and 50 $\mu$Hz, respectively.

Additional parameters linked to the quality of the stellar targets and of the acquired light curves are added: \textit{Kepler} magnitude values $K_p$ from the \textit{Kepler} input catalog \citep{2011AJ....142..112B}, length of the light curves (in days), bad quarter flag, number of bad quarters in the light curve, and finally, the start time and end time of the light curve.

For each star, a total of 159 input parameters are used to feed the ROOSTER pipeline. The full list of parameters is given in Table.~\ref{table:summary_param} in appendix~\ref{appendix:input_parameters}.

\section{The ROOSTER pipeline \label{section:pipeline}}

ROOSTER uses three random forest classifiers\footnote{The random forests classifiers are implemented with the Python \texttt{scikit-learn} package \citep{scikit-learn}.}: \textit{RotClass}, which selects stars that have a rotation signal; \textit{PollFlag}, which flags the Type-1 CP/CB candidates; and \textit{PeriodSel}, which allows ROOSTER to choose the final rotation period between the three different filters and the three analysis methods. Due to its ability to take into account a large number of parameters, its flexibility, and its adaptability to the training set, the random forest algorithm constitutes a clear improvement from the former method consisting in defining thresholds on a small number of key parameters.

\begin{figure}[ht]
    \centering
    \includegraphics[width=.49\textwidth]{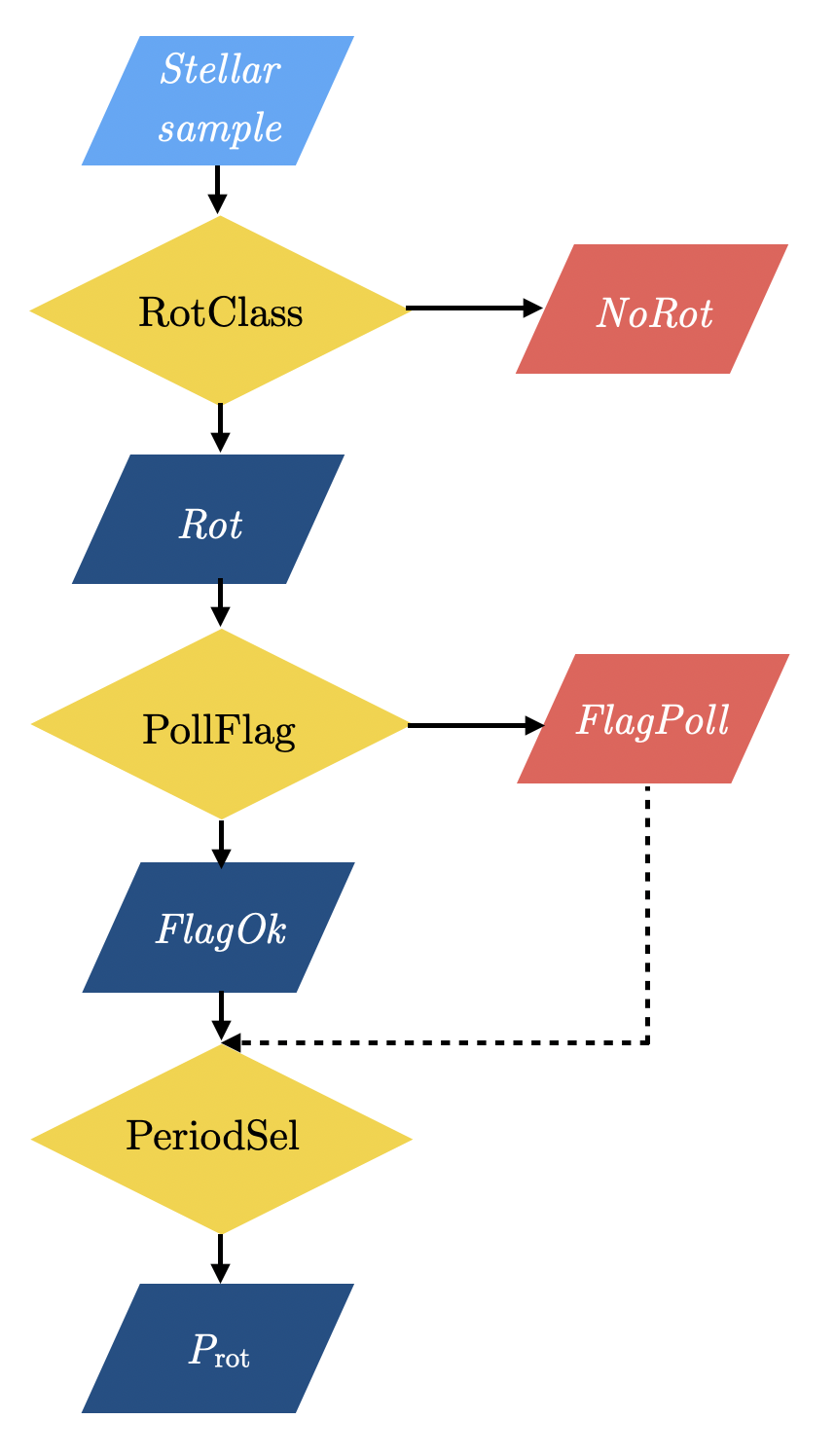}
    \caption{Structure of the ROOSTER pipeline. The yellow diamonds outline the action of each classifier used by the machine learning algorithm. The stars of the sample (lightblue parallelogram) are anlaysed by the pipeline. The red and dark blue parallelograms emphasis the label attributed by the ROOSTER classifiers to each star of the sample. First, stars with rotational modulation are selected by \textit{RotClass} and the label \textit{Rot} or \textit{NoRot} is given. \textit{NoRot} stars are not considered in the subsequent analysis. The \textit{PollFlag} classifier flags Type 1 CP/CB candidates (labelled \textit{FlagPoll}). The remainder of the stars receive the \textit{FlagOk} label. Finally, the \textit{PeriodSel} classifier selects the rotation period.}
    \label{fig:schema}
\end{figure}

The three classifiers are trained with the same 159 input parameters per star as defined in the previous section, but each classifier considers different training sets and gives different output labels. Depending on the task the classifier has to achieve, each of them may give different importance to the input parameters. The flow diagram of ROOSTER is shown in Fig.~\ref{fig:schema}.

\subsection{Description}
\begin{sidewaysfigure*}
    \includegraphics[width = 1. \textwidth]{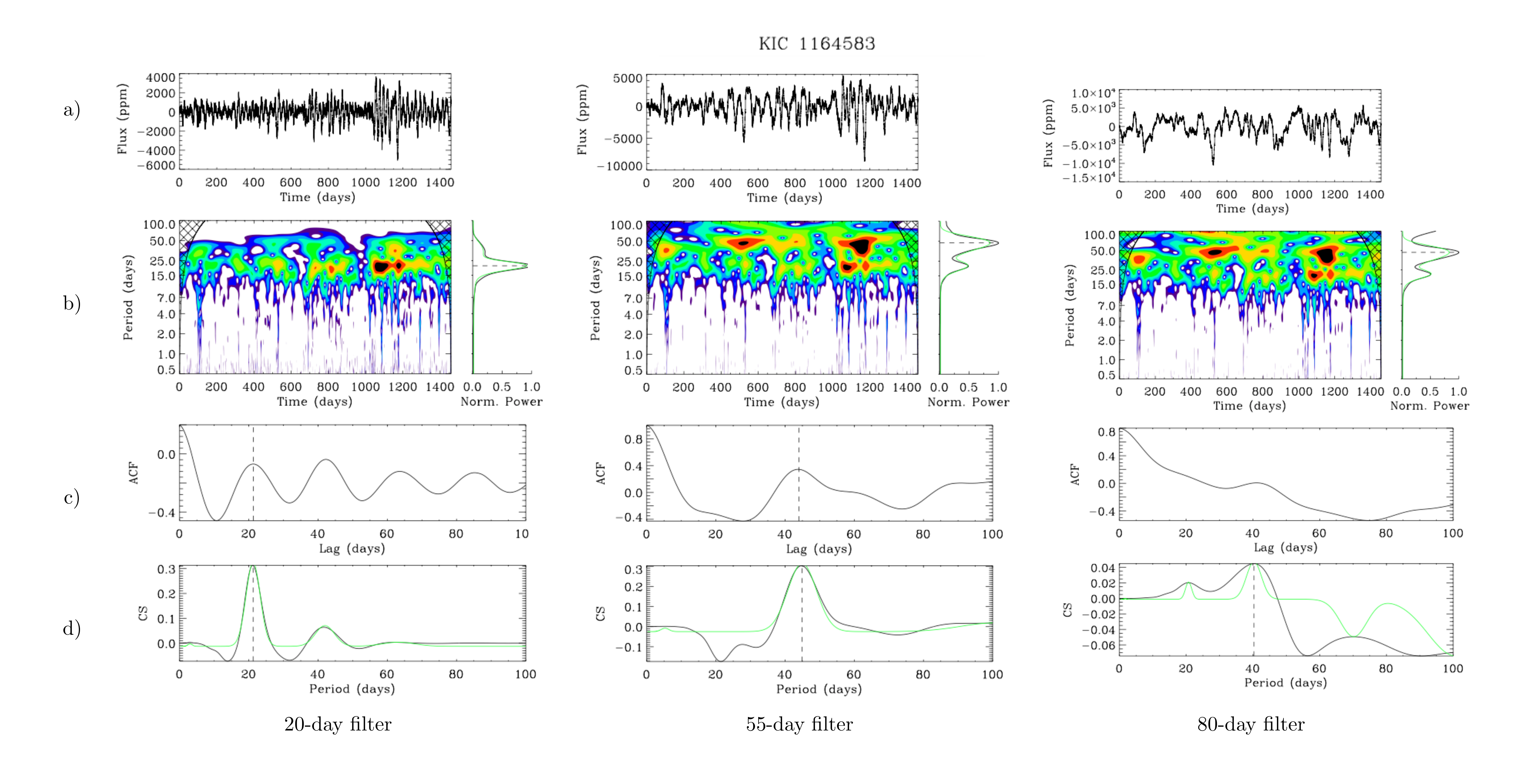}
    \caption{Comparison between the light curves obtained with the three high pass filters for KIC~1164583 (from left to right, 20, 55, and 80-day filter respectively). (a) KEPSEISMIC light curve. (b) Wavelet power spectrum (left) and global wavelet power spectrum (right). The grid represents the cone of influence of the reliable periodicity. High-amplitude power are highlighted by black and red whereas green and blue denote low-amplitude power. (c) Auto-Correlation Function. (d) Composite spectrum (black) and corresponding best fit with multiple Gaussian profiles (green). In all panels, the dashed line shows the position of the extracted rotation period.}
    \label{fig:filters}
\end{sidewaysfigure*}

The first ROOSTER classifier, \textit{RotClass}, separates stars with rotational modulations (labelled \textit{Rot} stars) from other stars 
(labelled \textit{NoRot} stars). These \textit{NoRot} stars are rejected and are no longer considered in the rest of the analysis (although some of those stars may later still be flagged for visual checks). 
An example of \textit{Rot} star, KIC~1164583, is presented in Fig.~\ref{fig:filters}. In this figure, the light curves and the output of the three different rotation period retrieval methods (WPS/GWPS, ACF, and CS) are shown.


In general, the risk of confusion by the ROOSTER pipeline between a \textit{Rot} and a \textit{NoRot} star is a consequence of two main issues: it can be difficult to reliably detect low-amplitude rotational modulations and \textit{Kepler} instrumental artifacts may introduce additional periodic modulations in the light curve. 

\begin{figure}[ht!]
    \includegraphics[width=0.5\textwidth]{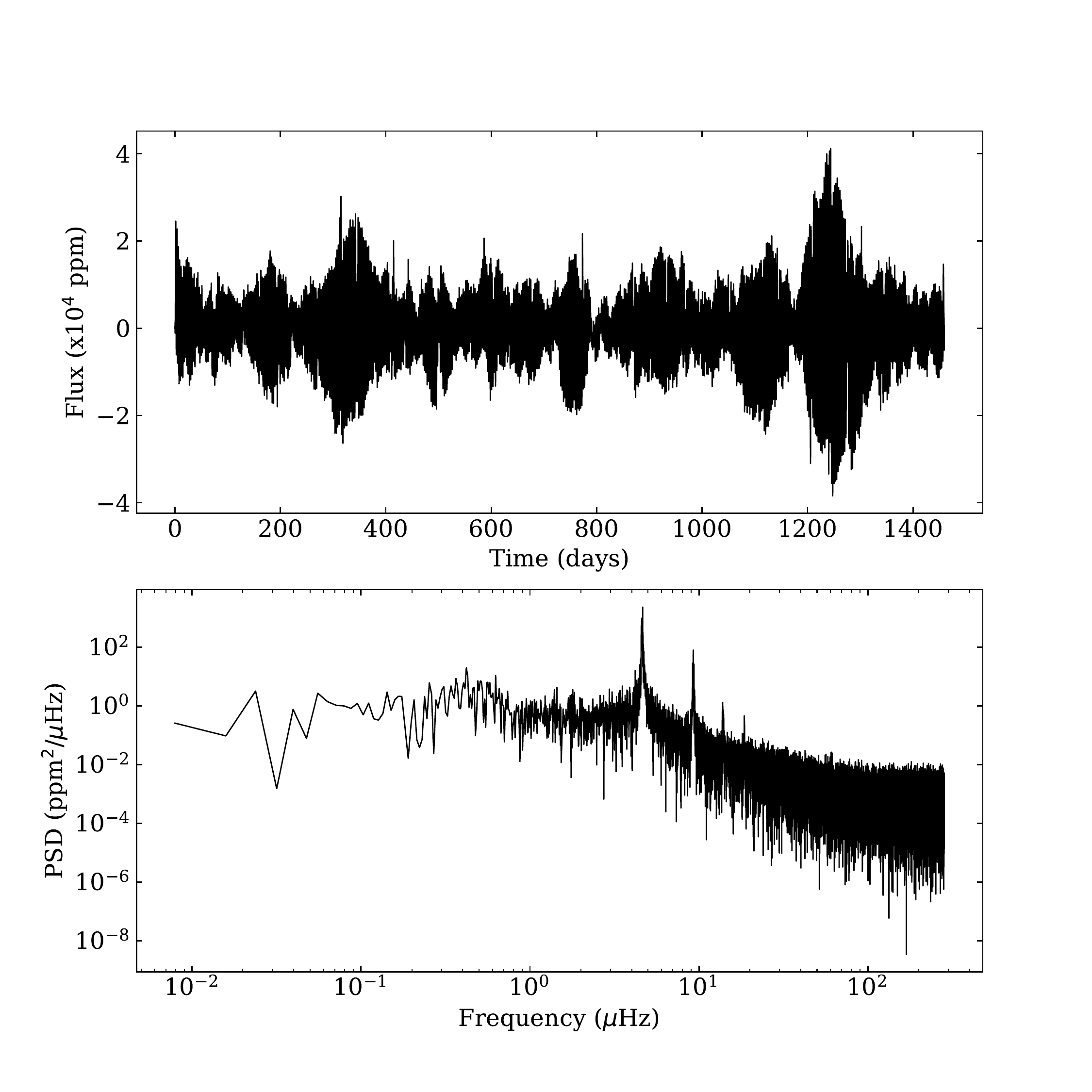}
    \caption{PSD of the Type 1 CP/CB candidate KIC~2283703. The light curve is characterised by high-amplitude flux variation and beating patterns over time, while the PSD exhibits a large number of high-amplitude peaks.}
    \label{fig:psd_cpcb}
\end{figure}

A small fraction of \textit{Rot} stars (generally stars with $P_\mathrm{rot} < 7$ days) may be close-in binaries or classical pulsator stars candidates. As mentioned above, the detected signal for this type of targets may not be consistent with rotation of single stars. The \textit{PollFlag} classifier is trained to identify these CP/CP candidates and flag them as \textit{FlagPoll}, while the rest of the stars are labelled \textit{FlagOk}.
The light curve and PSD of the \textit{FlagPoll} candidate, KIC~2283703, is shown as an example in Fig.~\ref{fig:psd_cpcb}. Its Type 1 CP/CB character can be recognised due to the large amplitude of the brightness variations, the beating pattern, and the large number of high-amplitude peaks in the PSD.


Finally, the third classifier, \textit{PeriodSel}, is trained with the same 159 parameters defined in Sect.~\ref{section:data_sample}. This time, the labels of the training set correspond to the nine possible rotation periods ($\PACF$, $\PCS$, $\PGWPS$ in the 20, 55, and 80-day filters). The goal is to select the most reliable period among those nine estimates, which may differ. For example, Fig.~\ref{fig:filters} illustrates the impact of the high-pass filter when producing the light curves of KIC~1164583. The 55- and 80-day filters indicate a 47-day period, while the period detected from the 20-day filter is 21 days. The comparison between the three WPS shows that the 47-day signal has been filtered out by the 20-day filter. Note also that the WPS shows power at longer periods which is another evidence for a longer period being filtered. The 47-day GWPS period of the 55-day filter is finally chosen as the retrieved $P_\mathrm{rot}$. \textit{PeriodSel} selects one of the nine estimates as the retrieved rotation period, $P_\mathrm{rot, ML}$. 
We emphasise that the transfer function of a given filter slowly approaches zero at twice the cutoff period, i.e. it is possible to retrieve rotation periods longer than the cutoff period of the filter applied to the light curve.

\subsection{Training \label{section:KM}}

The approach followed allows us to show the robustness of ROOSTER. In particular, ROOSTER must be weakly dependent to the exact composition of the training set. 
This can be only achieved if the training-set parameter space distribution is similar to the test-set parameter space distribution. 
Hence, we decide to perform a training-loop process where the three classifiers are trained one hundred times. 
Each time, the full working sample is randomly divided in a training set (containing 75\% of the stars) and a test set (containing 25\% of the stars). The number of times each star is drawn in the test set over the one hundred training repetitions follows a binomial law with probability $p=0.25$ and number of trials $n=100$.

In total, the target sample of this study includes 23,431 targets from S19: 21,707 of them have all the parameters from the rotation pipeline (referred as the \textit{RotClass} sample), while 1,724 have missing parameters (referred as the \textit{MissingParam} sample, mainly due to missing ACF values). S19 gives measured $P_\text{rot}$ for 14,936 stars: 14,562 for the \textit{RotClass} sample and 374 stars for the \textit{MissingParam} sample. These stars are labelled \textit{Rot} and are then used with the \textit{PeriodSel} classifier (called the \textit{PeriodSel} sample). 
A total of 8,495 stars do not have a reliable $P_\text{rot}$ in S19.
Hence, this S19 sample is ideal to train and evaluate the performance of ROOSTER. 
The \textit{MissingParam} sample is not considered at this stage but will be used later in Sect.~\ref{subsection:missing_param}.
The working \textit{PollFlag} sample is constituted of 630 stars: 315 Type 1 CP/CB candidates completed at each training with 315 other stars randomly chosen from the \textit{PeriodSel} sample.
The composition of the different samples is summarised in Table~\ref{tab:samples}.

\begin{table*}[h]
    \caption{Composition of the different subsamples used for the study and scores of the different classifiers. The 315 non-candidate stars used in the \textit{PollFlag} sample are randomly chosen in the \textit{PeriodSel} sample at each training. For \textit{RotClass} and \textit{PeriodSel}, the metrics chosen to express the score is the accuracy while it is the sensitivity for \textit{PollFlag}. When two scores are given, the first and the second one corresponds to the score before and after visual inspection, respectively. ($^*$) There is only 4 Type-1 CP/CB candidates in this sample, among which, only 2 have been correctly classified yielding 50\% probability.}
    \centering
    \begin{tabular}{c|cccc}
    \hline\hline 
& \textit{RotClass} sample & \textit{PollFlag} sample & \textit{PeriodSel} sample & \textit{MissingParam} sample \\
\hline
Number of input stars & 21,707 & \makecell{315 Type 1 CP/CB candidates \\ + 315 non-candidates} & 14,562 & 1,724 \\
\hline
\textit{RotClass} accuracy  (\%)  & 94.2/97.6 & - & -  & 77.4 \\
\textit{PollFlag} sensitivity (\%) & - & 99.4 & - & 50 $^*$ \\
\textit{PeriodSel} accuracy (\%) & - & - & 95.3/99.5 & 79.4 \\
global accuracy (\%) & - & - & 92.1/96.9 & - \\

    \hline
    \end{tabular}

    \label{tab:samples}
\end{table*}



A RF classifier not only gives a label to each classified star, its forest structure also allows obtaining the proportion of trees attributing a given class to each considered star. Indeed, the final classification of the target corresponds to the label that is selected by the majority of the trees. We can then define the classification ratio as the number of trees giving the output label over the total number of trees of the forest. This ratio is useful to estimate the reliability of the machine-learning output for each star. 

Since we perform a training loop with random repetitions, we are also able to compute the mean classification ratio for each star. This mean classification ratio is computed as the mean of all the classification ratios found each time a star has been drawn in the test set.

\section{Results \label{sec:results}}

In what follows, the \textit{accuracy} term is defined as the agreement of any label given by a ROOSTER classifier  and the S19 catalog taken as a reference. 

From the 21,707 stars of the \textit{RotClass} sample, the mean classification ratio gives the right label (\textit{Rot} or \textit{NoRot}) for 20,443 stars, which yields an overall accuracy of $\sim$ 94.2\%. We find that 1,707 stars have been misclassified at least once during the 100 runs (7.9\%) while 1259 stars have been misclassified in half of the repetitions of the training they were part of the test set (5.8\%). Finally, 884 stars have been systematically misclassified (4.1\%). The accuracy of \textit{RotClass} is already valuable but can be improved with a limited amount of visual checks. By performing visual inspections, we verify that the highest number of misclassifications occur within the mean classification ratio between 0.4 and 0.8. For this reason, stars within these mean classification ratios should be part of the visual verification for future target samples. In the case of the target sample of the current study, 2230 stars (10.3\%) are within this interval. One of the reasons for the misclassifications is the presence of instrumental modulations that can be mistaken by rotation periods.




Concerning \textit{PollFlag}, the goal is to obtain a high sensitivity to real Type-1 CP/CB candidates, while keeping the fraction of misclassifications low. The sensitivity is here defined as the fraction of Type 1 CP/CB candidates correctly identified by \textit{PollFlag}. Thus, the sensitivity does not take into account the fraction of non-Type 1 CP/CB candidates correctly identified.
In this case, it is a more relevant estimator than the accuracy, as the fraction of Type-1 CP/CB candidates in the sample is small. During the training loop, approximately 2.8\% of the considered non-Type-1 CP/CB candidates have been misclassified by \textit{PollFlag} as Type-1 CP/CB candidates. If \textit{PollFlag} had been applied blindly on the full \textit{PeriodSel} sample, this means that approximately 400 non-Type-1 CP/CB candidates would have been wrongly flagged. However, considering the mean classification ratio, only two Type-1 CP/CB candidates have been misclassified, which yields a \textit{PollFlag} sensitivity of $\sim$ 99.4\%. Looking at the detailed results during the training loop, 6 Type-1 CP/CB candidates are misclassified at least once (1.9\%), 4 are misclassified in half of the repetitions (1.3\%), and none is systematically misclassified.  


Finally, the mean classification ratio of the \textit{PeriodSel} classifier gives us an accuracy of $\sim$ 86.8\%, i.e. the label corresponding to the rotation period from S19 is attributed to 12,638 stars over 14,562 in the \textit{PeriodSel}. 
Nevertheless, a significant fraction of the labels chosen by \textit{PeriodSel} contains a $P_\mathrm{rot}$ which is within 10\% of the one in S19. For this reason, we adopt the agreement within 10\% as the \textit{true accuracy}, $\mathrm{TA}$, of \textit{PeriodSel} in the following way:

\begin{equation}
    \mathrm{TA} = \frac{\mathrm{Card}  \, (E_\mathrm{good})}{\mathrm{Card} \, (E)} \times 100 \;,
\end{equation}
with $E$ being the considered ensemble of stars and $E_\mathrm{good}$ the subsample from $E$ where \textit{PeriodSel} retrieves a $P_\mathrm{rot, ML}$ matching $P_\mathrm{rot, S19}$ within 10\%. 

\begin{figure}[ht]
    \includegraphics[width=0.5\textwidth]{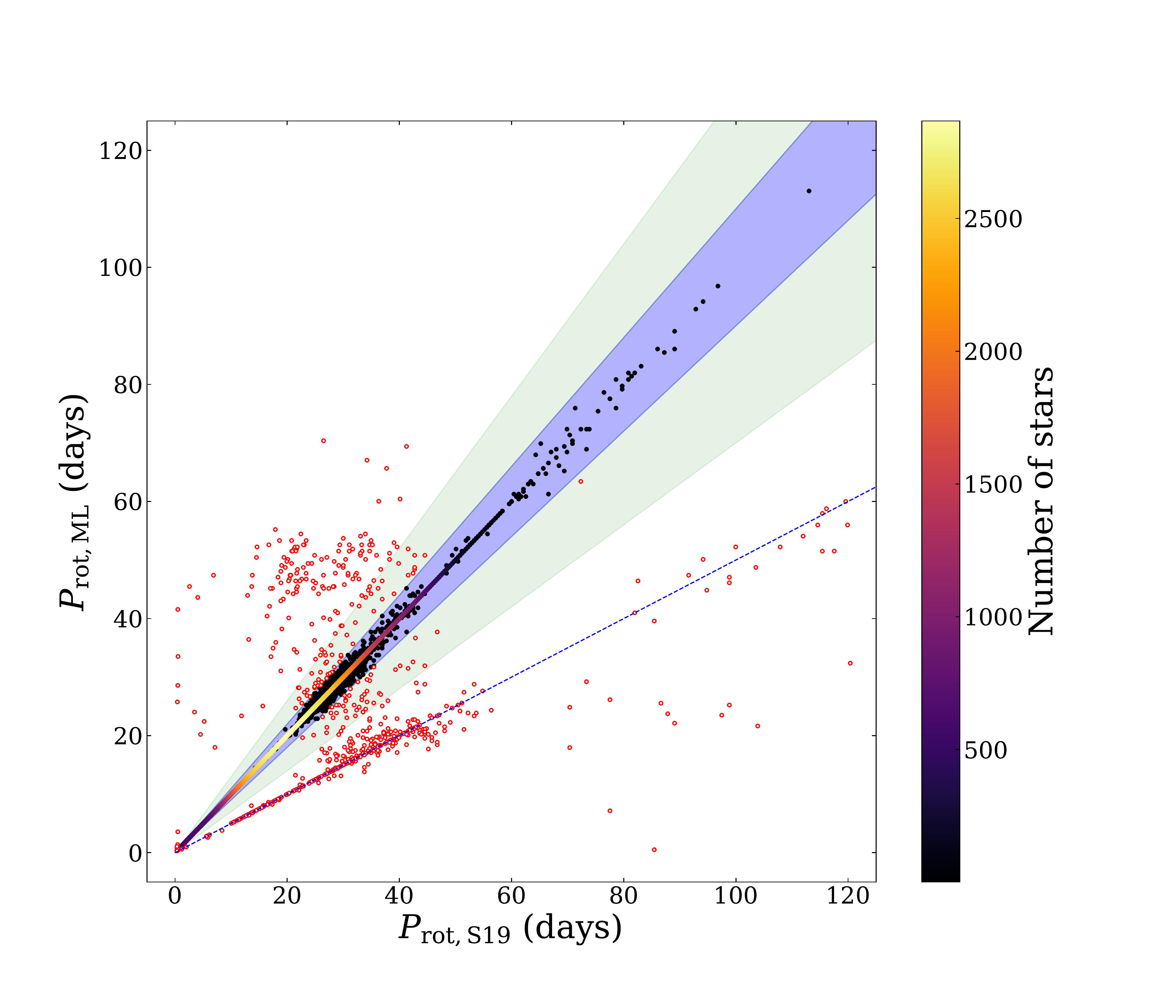}
    \caption{Comparison between $P_\mathrm{rot, ML}$ and the values given in S19 for the stars of the \textit{PeriodSel} sample. The shaded blue and green areas highlight the 10 and 30\% agreement regions, respectively. The number of stars inside this area is color coded. The stars outside the shaded area are represented by red dots. The dashed blue line corresponds 2:1 line.}
    \label{fig:true_accuracy}
\end{figure}

We find that \textit{PeriodSel} selected an incorrect period for 830 stars at least once (5.7\%), for 693 (4.8\%) in half of the realisations and systematically for 569 (3.9\%). In terms of the mean classification ratio, a correct rotation period $P_\mathrm{rot, ML}$ is retrieved for 13,871 stars. Hence, the true accuracy is $\sim$ 95.3\%. Fig.~\ref{fig:true_accuracy} shows the comparison between $P_\mathrm{rot, ML}$ and the reference values from S19. From the 1,924 stars for which the label attributed by ROOSTER was not the same as in S19, 1,233 are within the shaded blue area of the figure. For the remaining 4.7\% of the stars, we observe three cases:
\begin{enumerate}
    \item 103 periods retrieved by ROOSTER lie within a tolerance threshold that is slightly over 10\% (see the shaded green area of Fig.~\ref{fig:true_accuracy}). This could be due to a precision difference between $\PGWPS$, $\PACF$, and $\PCS$ or to the filter that was selected to choose the value of the rotation period.
    \item The machine learning mistakes instrumental modulations for real rotation signals when such signatures appear in the light curves. These instrumental modulations usually have periods between 38 and 50 days. 99 ROOSTER rotation periods lie both in this [38--50] days regions and outside of the 10\% agreement region. 
    \item Depending on the rotation signature, some of the nine period estimates correspond to the second harmonic of the correct rotation period. \textit{PeriodSel} privileges this harmonic over the fundamental period for 381 stars of the \textit{PeriodSel} sample. 
\end{enumerate}

\subsection{Weight of the considered parameters}

\begin{figure*}
    \includegraphics[width=1.\textwidth]{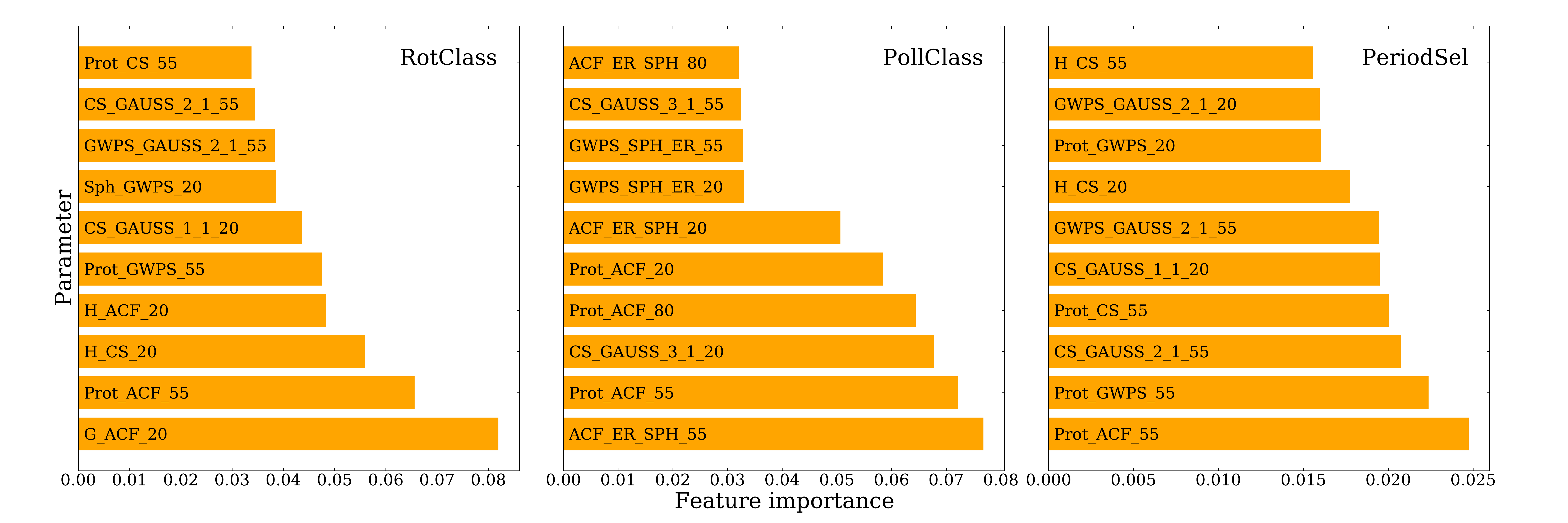}
   \caption{Importance of the 10 most important parameters for the three classifiers of the pipeline. The full list of ROOSTER input parameters is given in Appendix~\ref{appendix:input_parameters}.}
    \label{fig:parameter_importance}
\end{figure*}

Here we discuss briefly  the importance given by the three classifiers to the input parameters. For each classifier, Fig.~\ref{fig:parameter_importance} presents the weights of the 10 most important parameters. 
The rotation periods provided by the ACF and GWPS for the 55-day filter are in the top five parameters of \textit{RotClass}.
\textit{PollFlag} favours the ACF parameters, which is coherent as the ACF shows a specific pattern for CP/CB candidates. We note that \textit{PollFlag} is the only classifier among the three that also prioritises errors on \sph\ determination to perform the classification.
For \textit{PeriodSel}, three of the most important parameters are the rotation period values computed from the 55-day filtered light curves. The parameters of the Gaussian functions fitted to the GWPS and CS also have a strong influence over the classification. The remainder of the most important parameters are the $\HACF$, $\GACF$, and $\HCS$ from the 20-day filtered light curves, which highlights the importance of those control parameters when considering the reliability of a potential rotation signal.   

Finally, one could be concerned by the effect of systematic errors on $T_\mathrm{eff}$ and $\log g$ parameters. However, because of the small weight of these two parameters, their influence on our classifiers accuracy is marginal. Therefore, the impact on our methodology of any systematic error in both parameters is negligible. A similar conclusion will be reached in Sect.~\ref{sect:discussion_simulation} where we tested our algorithm using simulated data by \citet{2015MNRAS.450.3211A}.   

\subsection{Strategy for stars with missing parameters \label{subsection:missing_param}}

The rotation pipeline does not always provide a value for all the parameters needed for the random forest classification. The majority of the missing parameters are related to the ACF for the 55- and 80-day filters. In general this happens because there is no clear peak in the ACF.
However, to perform the classification, an input value must be provided for each of the 159 parameters. To test ROOSTER on the stars with missing parameters, we choose to train our three classifiers without the values for $\PACF$, $S_\mathrm{ph, ACF}$ and the respective uncertainty for the three filters. However, we decide to keep $\HACF$ and $\GACF$ as input parameters. If for a given star in a given filter, $\PACF$ is not determined, $\HACF$ and $\GACF$ will be set to zero. This is logical in the sense that the value for $\GACF$ and $\HACF$ indicates whether the corresponding peak is significant. 


Like before, \textit{RotClass} is trained with the \textit{RotClass} sample. Only 1,334 of the stars of the \textit{MissingParam} sample are correctly classified (accuracy of 77.4\% compared to the accuracy of 94.2\% reached with the \textit{RotClass} sample). This is not a surprise in the sense that when the ACF does not yield a period estimate, it suggests that the modulation in the light curve may not be clear enough. This assumption is confirmed when comparing the fraction of stars with rotation signal in the \textit{RotClass} sample and in the \textit{MissingParam} sample: 67.1\% of the stars in the \textit{RotClass} sample are labelled \textit{Rot} in S19 while only 24\% of the \textit{MissingParam} sample are labelled \textit{Rot} in S19. Thus, the two distributions are very different and this impacts the outcomes of the classifier. 

The accuracy loss is similar for \textit{PeriodSel}, with only 79.4\% of rotation period retrieved within 10\%. 
Among the \textit{MissingParam} sample, there are only four Type 1 CP/CP candidates, which make it difficult to estimate the sensitivity for this sample. Two of those candidates are correctly flagged (50\%).  

In order to optimise the efficiency of the pipeline, it is thus preferable to be able to compute upstream a full set of parameters for a maximal number of stars and to avoid having to deal with a too large \textit{MissingParam} sample.  

\section{Discussion \label{section:discussion}}

\begin{figure*}[h!]
    \includegraphics[width=1.\textwidth]{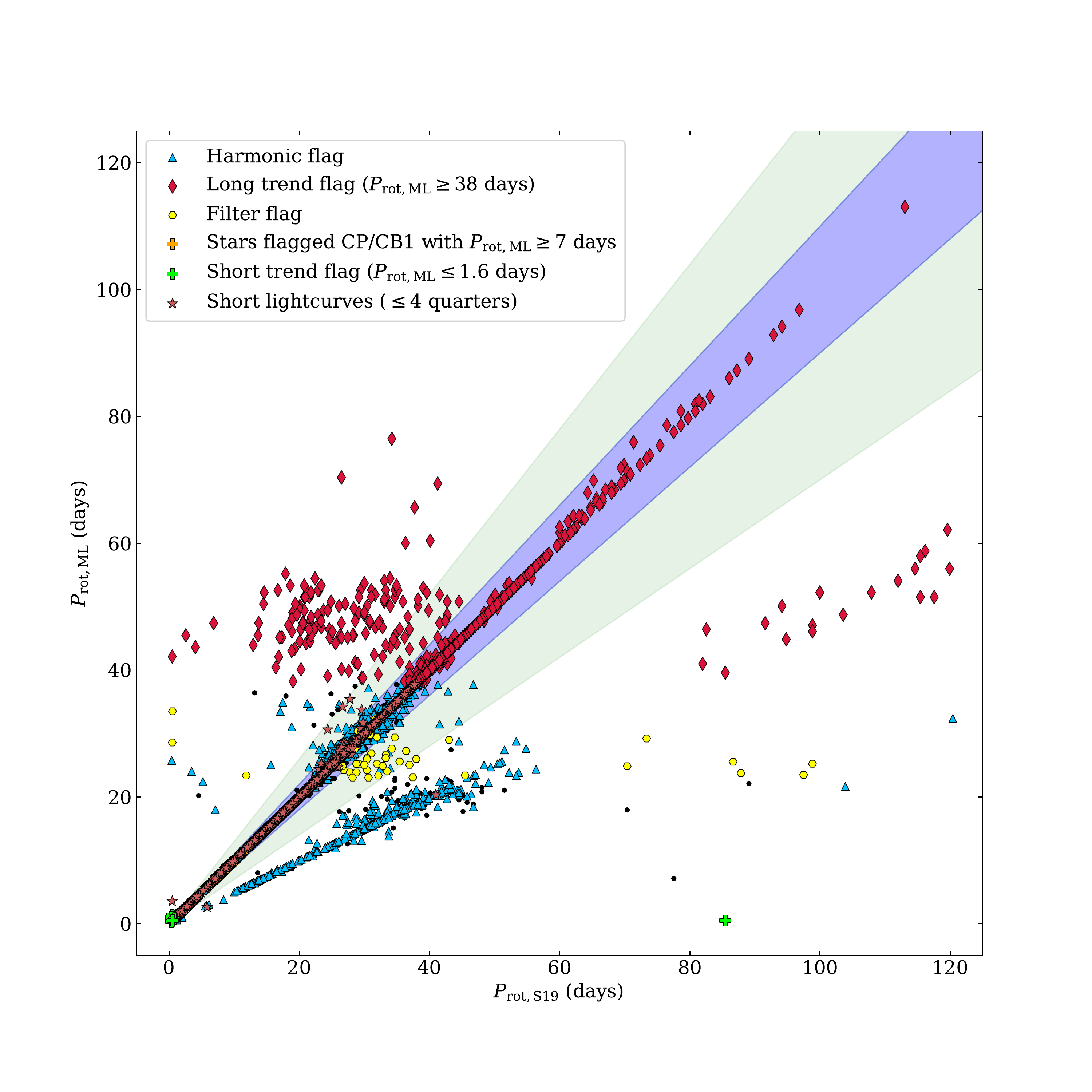}
    \caption{Comparison between the ML retrieved period and the value given in S19. The 10\% tolerance region is represented in blue, while the large tolerance zone (30\%) is shown in green. Stars that are flagged for visual checks are highlighted: yellow circle show the stars selected with the filter condition, red diamonds those by the long trend condition, cyan triangle those selected by harmonic search condition, brown stars those selected with the short light curve condition, green cross those selected by the short trend condition, and orange cross the stars flagged as Type 1 CP/CB with $P_\mathrm{rot,ML}>7$ days.}
    \label{fig:summary_figure}
\end{figure*}

The RF classification with ROOSTER represents a clear improvement compared to the automatic selection method based on control parameters performed in S19 (see Appendix~\ref{appendix:our_methodology}). 
However, for a small fraction of the sample, the ROOSTER results differ from those in S19. Nevertheless, the discrepancies represent well-known issues that we will try to flag.  


Once we have obtained the rotation periods from the ROOSTER pipeline, we define a list of criteria so that we can flag stars that require visual checks and hence improve the accuracy of the final retrieval of extracted rotation periods.

\begin{enumerate}

\item First, we identify stars for which the proper filter is not selected. We expect $P_\mathrm{rot,ML}$ chosen by \textit{PeriodSel} to be coherent with the filter from which they were extracted. This means that, for $P_\mathrm{rot,ML} < 23$ days, $P_\mathrm{rot,ML}$ should have been extracted from the 20-day filtered light curve, for $23$ days $< P_\mathrm{rot,ML} < 60$ days from the 55-day filtered light curve, for $P_\mathrm{rot,ML} > 60$ days from the 80-day filtered light curve. If $P_\mathrm{rot,ML}$ agrees within 15\% with the $\PGWPS$ of the proper filter, we replace $P_\mathrm{rot,ML}$ by this value. Otherwise, the star is flagged for visual check. This change is motivated by the goal of retrieving the most accurate \sph, which is generally obtained considering a light curve with a high-pass filter coherent with $P_\mathrm{rot,ML}$. 

\item As shown in Fig.~\ref{fig:true_accuracy}, a significant number of stars lie on the 1:2 line, highlighting an issue with the harmonics of $P_\mathrm{rot}$, which can partially be overcome. We flag targets for visual inspection if at least one of the other \prot\ estimates (nine in total) is within $15\%$ of the double of $P_\mathrm{rot,ML}$. 
We notice that this harmonics issue is generally linked to \prot\ yielded by the GWPS fitting method that selects a higher-order harmonic instead of the first one as it is more often done by the ACF methodology. In order to reduce the impact of this problem, it would be necessary to perform a comprehensive study of the fitted GWPS harmonic’s pattern. This analysis is out of the scope of the current paper but will be the issue of a future upgrade of the ROOSTER pipeline. 


\item The ROOSTER pipeline might also attribute the wrong rotation period because of instrumental modulation in \textit{Kepler} data. The 55 and 80 day filtered light curves are the most affected by instrumental modulations. For these cases, $P_\mathrm{rot,ML}$ is usually between 38 and 50 days, {but can be longer}.
Those instrumental modulations can be identified through visual inspection of the light curves and of the control figures of the rotation code. For this reason, we flag all stars with $P_\mathrm{rot,ML}>38$ days. 
Instrumental modulations can correspond to, for example: sudden bursts in flux; abnormal flux variations which only happen every four \textit{Kepler} Quarters, i.e. every \textit{Kepler} year; and flux variations that are present throughout the light curve with a main periodicity of one \textit{Kepler} year, while in the rotation diagnostics used here one can identify its harmonics and the resulting filtered signals. We note that the 80-day filtered KEPSEISMIC light curves are naturally the most affected by instrumental modulations. However, the signature in the 20-day and 55-day filtered light curves may also have an instrumental origin. This is also true for other data products in the literature. Note that, avoiding instrumental modulations is one of our objectives and one of the reasons to use three KEPSEISMIC light curves (Sect \ref{subsection:data_calibration}) and perform complementary visual checks (see also discussion in S19).



\item Stars with $P_\mathrm{rot,ML} < 1.6$ days are also flagged to ensure that the detected signal is not CP/CB-like. Observationally, synchronized binaries correspond to targets with detected period shorter than seven days \citep{Simonian2019}. 

\item All the Type~1 CP/CB candidates in S19 have a period shorter than seven days. We therefore flag for visual inspection Type-1 CP/CB candidates detected by \textit{PollFlag} when $P_\mathrm{rot,ML}>7$ days. 

\item Stars for which the length of data or the presence of gaps might not be enough to cover the range of rotation periods expected for the studied stars are also visually inspected. The flagging criterion can be based on the total length of observations (given in days or quarters) and on the continuity of the light curve. In this work we have used a criterion in quarters. Stars were flagged if their light curve was shorter than five quarters. 

\item The last group of visual checks correspond to stars are those for which the mean classification ration of \textit{RotClass} is between 0.4 and 0.8, as already stated in Sect.~\ref{sec:results}.

\end{enumerate}

The different types of stars flagged for visual inspection are summarised in Fig.~\ref{fig:summary_figure}. 

The six conditions listed above result in flagging for visual inspection 4,606 stars (31.6\% of the \textit{PeriodSel} sample). Among those 4,606 stars, 628 will be corrected after the visual inspection. Finally, for 27 stars, we prefer the value given by the machine learning over the S19 value. 
After this step, we obtain a 99.5\% accuracy for the retrieval of $P_\mathrm{rot,ML}$. It should be noted that among the 63 remaining stars outside the 10\% agreement between ROOSTER and S19, 24 of them lie in a large tolerance zone of 30\% of the good rotation period. 
As mentioned in Sect.~\ref{sec:results}, stars with mean classification ratio between 0.4 and 0.8 are visually checked. That means that some of the stars classified \textit{NoRot} by \textit{RotClass} are also visually checked. 
The total number of visual checks to perform in the scheme we just described is then 5,461 (25.2\% of the \textit{RotClass} sample), which represent a significant improvement compared to S19 (60\% of the sample was visually checked).

We are able to compute a global accuracy score for both \textit{RotClass} and \textit{PeriodSel} classifiers. Among the 14,562 stars of the \textit{PeriodSel} sample, both \textit{RotClass} and \textit{PeriodSel} made correct predictions for 13,413 stars (92.1 \% of the sample), i.e., \textit{RotClass} correctly labelled them \textit{Rot} and $P_\mathrm{rot,ML}$ agrees within 10\% of the $P_\mathrm{rot}$ given in S19. After the visual checks, this accuracy is raised to 96.9\%. The accuracy scores before and after visual inspection are summarised in Table~\ref{tab:samples}.

Note that it could be possible to reduce the global amount of visual checks. For example, 683 stars are flagged only because their light curve was shorter than five quarters. It represents 4.7\% of the \textit{PeriodSel} sample. Among those 683 stars, only 7 of them are actually corrected after the visual inspection. The visual check of short light curves could therefore be avoided without a significant loss of accuracy. However, concerning the methodology described in this paper, we decided to clearly point out every risk of confusion that could occur in ROOSTER.

\subsection{Period retrieval in short light curves}

\begin{figure*}[h!]
    \includegraphics[width=1. \textwidth]{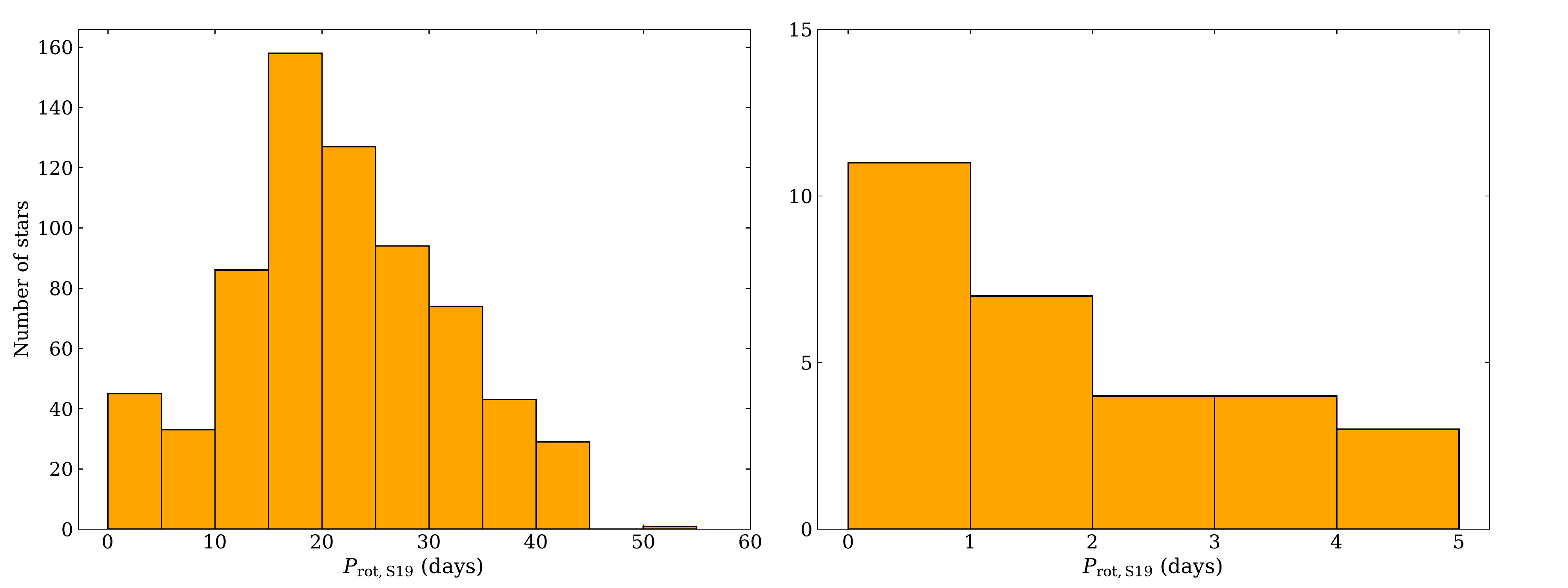}
    \caption{
                \textit{Left}: \prot\ distribution for stars with light curves within 35 and 150 days. \textit{Right}: Same as left panel, but for stars with light curves within 30 and 35 days.
            }
    \label{fig:short_lc_hist}
\end{figure*}

The ROOSTER accuracy on short light curves may yield interesting hints about the performance of our method on a dataset constituted by stars observed by TESS or K2. 
Our dataset contains 29 light curves with length within 30 and 35 days and 690 light curves with length within 35 and 150 days. \textit{PeriodSel} accuracy over those two subsamples is 96.5\% and 92.9\%, respectively. The \prot\ distribution for those two subsamples is presented in Fig.~\ref{fig:short_lc_hist}. All the stars with light curves length below 35 days present rotation periods below five days. The ability of our method to retrieve longer \prot\ in such short light curves is beyond the scope of this paper and will be discussed in details in a subsequent study.  

\subsection{Period retrieval with simulated data \label{sect:discussion_simulation}}

ROOSTER was also applied on the simulated data of the hare and hounds exercise in \citet{2015MNRAS.450.3211A}. The results of this analysis are presented in Appendix~\ref{appendix:benchmark_simulations}. For the purpose of this exercise, ROOSTER was trained considering only light curves from S19 and had no previous knowledge of the simulated data. 
We found that ROOSTER applied blindly performed better than any of the methods compared in \citep{2015MNRAS.450.3211A}. The ROOSTER accuracy scores are close to what we obtain on real data in this study. 
It is important to notice that in this analysis $T\mathrm{eff}$ and $\log g$ were not used in the training in order to have the same homogeneous set of parameters for the noise-free and noisy light curves.

\section{Conclusion \label{section:conclusion}}

In this paper, we have described ROOSTER, a machine learning analysis pipeline dedicated to obtain rotation periods with a small amount of visual verification. ROOSTER has been successfully applied to targets observed by the \textit{Kepler} main mission. The pipeline is built around three random forest classifiers, each one in charge of a dedicated task: detecting rotational modulations, flagging close binary and classical pulsator candidates, and selecting the correct rotation period, respectively with the classifiers \textit{RotClass}, \textit{PollFlag} and \textit{PeriodSel}.  

We applied ROOSTER to the K and M main-sequence dwarfs analysed in S19 \citep{2019ApJS..244...21S}.
We were able to detect rotation periods with an accuracy of 94.2\% for a sample of 21,707 stars (\textit{RotClass} sample). Within the \textit{RotClass} sample, we considered 14,562 stars with measured rotation periods (\textit{PeriodSel} sample). The automatic analysis yields a true accuracy of 95.3\% for \textit{PeriodSel}, i.e. an agreement of the rotation period from ROOSTER within 10\% of the reference value from S19. We noticed that the majority of the misclassified stars were due to known effects, in particular, instrumental modulations and confusion with a harmonic of the true rotation period. 31.6\% of the stars of the \textit{PeriodSel} sample are then flagged for visual checks which allows us to correct wrong values for 628 more stars and raises the \textit{PeriodSel} accuracy to 99.5\%.  Considering the results of the combined \textit{RotClass} and \textit{PeriodSel} steps on the \textit{PeriodSel} sample, the global accuracy of ROOSTER is estimated at 92.1\% without any visual check, and 96.9\% after visual inspection of 25.2 \% of the \textit{RotClass} sample (against 60\% in S19). By removing some input parameters from the ROOSTER training, we were also able to deal with stars with missing input parameters, at the price of a loss of accuracy. However, this accuracy loss is not only due to the diminution of the number of the parameters, but is also related to the intrinsic quality of the light curves for which the rotation pipeline was not able to provide a full set of parameters. 

It should be emphasised that the accuracy estimated here is only valid for the K and M dwarf stars studied in S19. Rotation signals for hotter stars (mid F to G) are more complex than for K and M. In order to extend our analysis to those stars, a significant sample of hotter stars should first be introduced in the training set. The different accuracy values should also be reassessed before proceeding to the analysis. 
The length of the \textit{Kepler} time series and the exquisite quality of the photometric measurements are important advantages for rotational analyses. To analyse light curves from K2 or TESS (which have a typical observational length  significantly shorter than those of \textit{Kepler}), first we need to verify whether the current training set is appropriate. If this is not the case, the classifiers must be trained with more suitable training sets (e.g by building a new training set with K2 or TESS stars). The selection of targets for visual inspection would probably be modified too.    

Nevertheless, the global framework of ROOSTER can be adapted the \textit{Kepler} F and G main-sequence solar type stars (Santos et al. in prep). It should also be possible to use this technique to analyse data from K2, TESS and, in a near future, PLATO \citep[PLAnetary Transit and Oscillations,][]{2014ExA....38..249R}. Such large-scale survey would be a great asset for gyrochronology models and our understanding of stellar spinning evolution in relation with age and spectral type.   

\begin{acknowledgements}
We thank Suzanne Aigrain and Joe Llama for providing us with the simulated data used in \cite{2015MNRAS.450.3211A}.
S.N.B., L.B. and R.A.G. acknowledge the support from PLATO and GOLF CNES grants. A.R.G.S. acknowledges the support from NASA under grant NNX17AF27G. S.M. acknowledges the support from the Spanish Ministry of Science and Innovation with the Ramon y Cajal fellowship number RYC-2015-17697. P.L.P. and S.M. acknowledge support from the Spanish Ministry of Science and Innovation with the grant number PID2019-107187GB-I00. This research has made use of the NASA Exoplanet Archive, which is operated by the California Institute of Technology, under contract with the National Aeronautics and Space Administration under the Exoplanet Exploration Program.    
\\
\textit{Software:} \texttt{Python} \citep{10.5555/1593511}, \texttt{numpy} \citep{numpy}, \texttt{pandas} \citep{reback2020pandas, mckinney-proc-scipy-2010}, \texttt{matplotlib} \citep{4160265}, \texttt{scikit-learn} \citep{scikit-learn}.
\\
The source code used to obtain the present results can be found at: \\ \texttt{gitlab.com/sybreton/pushkin} \\ \texttt{gitlab.com/sybreton/ml\_surface\_rotation\_paper}.
\end{acknowledgements}

%
\bibliographystyle{aa} 
\bibliography{biblio.bib} 
%

\appendix

\section{Detailed parameter extraction methodology \label{appendix:our_methodology}}

This appendix presents in detail of the methodology used to extract the $P_\mathrm{rot}$ candidates considered by ROOSTER.

The first method used to measure $P_\text{rot}$ is a time-period analysis using a wavelet decomposition \citep{1998BAMS...79...61T,10.1175/2007JTECHO511.1,2010A&A...511A..46M}. Wavelets of different periods, each one taken as the convolution between a sinusoidal and a Gaussian function (Morlet wavelet), are cross-correlated with the light curve to obtain the wavelet power spectrum (WPS, see left panel of b) in Figure 2). The WPS is projected over the period-axis to obtain the one-dimension Global Wavelet Power Spectrum (GWPS, see right panel of b) in Figure 2). Multiple Gaussian functions are fitted to the GWPS starting by the one with the highest amplitude. The fitted Gaussian function is removed and the next highest Gaussian peak is fitted in an iterative way until no peaks are above the noise level. Thus, the first period estimate, corresponding to the period of the highest fitted peak in the GWPS, is assigned as the rotation period recovered by this methodology: $\PGWPS$. The period uncertainty is taken as the half width at half maximum (HWHM) of the Gaussian function. Although this is a very conservative approach, it allows us to take into account variations due to differential rotation as part of the uncertainty. 

The second analysis method is the ACF of the light curve. The rotation period, $\PACF$, corresponds to the period of the highest peak in the ACF at a lag greater than zero. Two other parameters are computed: $\GACF$ and $\HACF$. $\GACF$ is the height of $\PACF$, while $\HACF$ is the mean difference between the height of $\PACF$ and the two local minima on both sides of $\PACF$ \citep[see an extended description in][]{2017A&A...605A.111C}.

The third method is the composite spectrum (CS). CS is the product between the normalized GWPS and ACF. This way, the peaks present in both GWPS and ACF (possibly related to rotation) are amplified while the signals appearing only in one of the two methods (for example due to instrumental effects that have a different manifestation in each analysis) are attenuated. Multiple Gaussian functions are fitted to the CS following the same iterative procedure as the one described for the GWPS. $\PCS$ is obtained as the period of the fitted Gaussian of highest amplitude and the uncertainty is its HWHM. The amplitude of this peak is $\HCS$.

Having three period estimates for each light curve ($\PGWPS$, $\PACF$, and $\PCS$), S19 computed the respective value for the photometric activity index \sph\  (i.e., one for each $P_\mathrm{rot}$ estimation). \sph\ is computed as defined by \citet{2014A&A...562A.124M}, being the standard deviation over light curve segments of $5\times P_\text{rot}$. The final \sph\ value we provide corresponds to the average \sph. \sph\ is corrected for the photon-shot noise following \citet{Jenkins2010}. For some targets this correction leads to negative \sph\ values. Most of the targets in this situation do not show rotational modulation. For those with rotational modulation and \sph\ < 0 (a few percent of the targets), S19 applied individually a different correction to the photon-shot noise, which is computed from the high-frequency noise component in the power density spectrum. We note that the \sph\ value has only a physical sense when rotational modulation is detected in the light curve and, thus, \prot\ is measured \citep[e.g.][Egeland et al. in prep]{2014A&A...562A.124M}. 

In S19, a first group of stars with reliable rotation periods is automatically selected when there is a good agreement between the different period estimates and the heights for $\PACF$ and $\PCS$ are larger than a given threshold \citep[see S19 for details; height thresholds adopted from][]{2017A&A...605A.111C}. For the remainder of the targets (about $60\%$), in order to decide whether the signal is related to rotational modulation and to select the correct \prot, S19 visually inspected the respective light curves, rotation diagnostics, and power spectra. During the visual inspection $40\%$ of the final selected \prot\ were recovered. There are different problems causing the measurement of a different \prot\ in each methodology. For example: small amplitudes of the rotational modulation, which translate into small values of \sph, $\HACF$, $\GACF$, and $\HCS$; presence of instrumental modulations; and strong harmonic of \prot. Instrumental modulations affect primarily the light curves obtained with the 55-day and 80-day filters. However, we note that depending on the \prot\ value, the correct period may not be recovered from the 20-day filter. Half of the rotation period can be wrongly retrieved as \prot\ (what we call strong harmonic) when the dominant spots producing the rotational signal are apart by $\sim180^\circ$ in longitude. In our methodology, the GWPS is the most sensitive to this issue, while the ACF is the least sensitive. The performance of the ACF in these cases is discussed in detail by \cite{2013ApJ...775L..11M,2014ApJS..211...24M}. In S19, the decision on  the filter used for the final \prot\ relies on two criteria: preserving the instrinsic rotational signal with the \sph\ value not affected by filtering, while minimizing the impact of possible instrumental effects which may not alter the period estimate but may affect \sph. Typically the 20-day filter is selected for $P_\text{rot}\leq23$ days, the 55-day filter is selected for $23<$ days $P_\text{rot}\leq 60$ days, and the 80-day filter is selected for $P_\text{rot}> 60$ days. For the rotation period estimate itself the priority is given to $\PGWPS$. The main reason for this choice is the conservative uncertainty for $\PGWPS$.

\section{Input parameters \label{appendix:input_parameters}}

The detailed set of 159 parameters used to train \textit{RotClass}, \textit{PollFlag} and \textit{PeriodSel} is presented in Table~\ref{table:summary_param}. GWPS\_GAUSS\_1\_i\_XX, GWPS\_GAUSS\_2\_i\_XX and GWPS\_GAUSS\_3\_i\_XX respectively correspond to the amplitude, the central period and the standard deviation of the $i^\mathrm{th}$ Gaussian fitted in the GWPS with the XX-day filter. The same naming convention has been used for CS\_GAUSS\_1\_i\_XX, CS\_GAUSS\_2\_i\_XX and CS\_GAUSS\_3\_i\_XX with the CS. 

\begin{table}[ht]
\caption{List of the 159 input parameters used to train the RF classifiers.
\label{table:summary_param}}
\centering
\begin{adjustbox}{width=.5\textwidth}
\tiny
\begin{tabular}{lll}
\hline\hline
     ACF\_ER\_SPH\_20 &      ACF\_ER\_SPH\_55 &      ACF\_ER\_SPH\_80 \\
        BAD\_Q\_FLAG &         CS\_CHIQ\_20 &         CS\_CHIQ\_55 \\
        CS\_CHIQ\_80 &    CS\_GAUSS\_1\_1\_20 &    CS\_GAUSS\_1\_1\_55 \\
   CS\_GAUSS\_1\_1\_80 &    CS\_GAUSS\_1\_2\_20 &    CS\_GAUSS\_1\_2\_55 \\
   CS\_GAUSS\_1\_2\_80 &    CS\_GAUSS\_1\_3\_20 &    CS\_GAUSS\_1\_3\_55 \\
   CS\_GAUSS\_1\_3\_80 &    CS\_GAUSS\_1\_4\_20 &    CS\_GAUSS\_1\_4\_55 \\
   CS\_GAUSS\_1\_4\_80 &    CS\_GAUSS\_1\_5\_20 &    CS\_GAUSS\_1\_5\_55 \\
   CS\_GAUSS\_1\_5\_80 &    CS\_GAUSS\_2\_1\_20 &    CS\_GAUSS\_2\_1\_55 \\
   CS\_GAUSS\_2\_1\_80 &    CS\_GAUSS\_2\_2\_20 &    CS\_GAUSS\_2\_2\_55 \\
   CS\_GAUSS\_2\_2\_80 &    CS\_GAUSS\_2\_3\_20 &    CS\_GAUSS\_2\_3\_55 \\
   CS\_GAUSS\_2\_3\_80 &    CS\_GAUSS\_2\_4\_20 &    CS\_GAUSS\_2\_4\_55 \\
   CS\_GAUSS\_2\_4\_80 &    CS\_GAUSS\_2\_5\_20 &    CS\_GAUSS\_2\_5\_55 \\
   CS\_GAUSS\_2\_5\_80 &    CS\_GAUSS\_3\_1\_20 &    CS\_GAUSS\_3\_1\_55 \\
   CS\_GAUSS\_3\_1\_80 &    CS\_GAUSS\_3\_2\_20 &    CS\_GAUSS\_3\_2\_55 \\
   CS\_GAUSS\_3\_2\_80 &    CS\_GAUSS\_3\_3\_20 &    CS\_GAUSS\_3\_3\_55 \\
   CS\_GAUSS\_3\_3\_80 &    CS\_GAUSS\_3\_4\_20 &    CS\_GAUSS\_3\_4\_55 \\
   CS\_GAUSS\_3\_4\_80 &    CS\_GAUSS\_3\_5\_20 &    CS\_GAUSS\_3\_5\_55 \\
   CS\_GAUSS\_3\_5\_80 &    CS\_NOISE\_20 &          CS\_NOISE\_55 \\
       CS\_NOISE\_80 &        CS\_N\_FIT\_20 &        CS\_N\_FIT\_55 \\
       CS\_N\_FIT\_80 &       CS\_SPH\_ER\_20 &       CS\_SPH\_ER\_55 \\
      CS\_SPH\_ER\_80 &           END\_TIME &               F\_07 \\
              F\_20 &               F\_50 &                F\_7 \\
      GWPS\_CHIQ\_20 &       GWPS\_CHIQ\_55 &       GWPS\_CHIQ\_80 \\
 GWPS\_GAUSS\_1\_1\_20 &  GWPS\_GAUSS\_1\_1\_55 &  GWPS\_GAUSS\_1\_1\_80 \\
 GWPS\_GAUSS\_1\_2\_20 &  GWPS\_GAUSS\_1\_2\_55 &  GWPS\_GAUSS\_1\_2\_80 \\
 GWPS\_GAUSS\_1\_3\_20 &  GWPS\_GAUSS\_1\_3\_55 &  GWPS\_GAUSS\_1\_3\_80 \\
 GWPS\_GAUSS\_1\_4\_20 &  GWPS\_GAUSS\_1\_4\_55 &  GWPS\_GAUSS\_1\_4\_80 \\
 GWPS\_GAUSS\_1\_5\_20 &  GWPS\_GAUSS\_1\_5\_55 &  GWPS\_GAUSS\_1\_5\_80 \\
 GWPS\_GAUSS\_2\_1\_20 &  GWPS\_GAUSS\_2\_1\_55 &  GWPS\_GAUSS\_2\_1\_80 \\
 GWPS\_GAUSS\_2\_2\_20 &  GWPS\_GAUSS\_2\_2\_55 &  GWPS\_GAUSS\_2\_2\_80 \\
 GWPS\_GAUSS\_2\_3\_20 &  GWPS\_GAUSS\_2\_3\_55 &  GWPS\_GAUSS\_2\_3\_80 \\
 GWPS\_GAUSS\_2\_4\_20 &  GWPS\_GAUSS\_2\_4\_55 &  GWPS\_GAUSS\_2\_4\_80 \\
 GWPS\_GAUSS\_2\_5\_20 &  GWPS\_GAUSS\_2\_5\_55 &  GWPS\_GAUSS\_2\_5\_80 \\
 GWPS\_GAUSS\_3\_1\_20 &  GWPS\_GAUSS\_3\_1\_55 &  GWPS\_GAUSS\_3\_1\_80 \\
 GWPS\_GAUSS\_3\_2\_20 &  GWPS\_GAUSS\_3\_2\_55 &  GWPS\_GAUSS\_3\_2\_80 \\
 GWPS\_GAUSS\_3\_3\_20 &  GWPS\_GAUSS\_3\_3\_55 &  GWPS\_GAUSS\_3\_3\_80 \\
 GWPS\_GAUSS\_3\_4\_20 &  GWPS\_GAUSS\_3\_4\_55 &  GWPS\_GAUSS\_3\_4\_80 \\
 GWPS\_GAUSS\_3\_5\_20 &  GWPS\_GAUSS\_3\_5\_55 &  GWPS\_GAUSS\_3\_5\_80 \\
     GWPS\_NOISE\_20 &      GWPS\_NOISE\_55 &      GWPS\_NOISE\_80 \\
     GWPS\_N\_FIT\_20 &      GWPS\_N\_FIT\_55 &      GWPS\_N\_FIT\_80 \\
    GWPS\_SPH\_ER\_20 &     GWPS\_SPH\_ER\_55 &     GWPS\_SPH\_ER\_80 \\
          G\_ACF\_20 &           G\_ACF\_55 &           G\_ACF\_80 \\
          H\_ACF\_20 &           H\_ACF\_55 &           H\_ACF\_80 \\
           H\_CS\_20 &            H\_CS\_55 &            H\_CS\_80 \\
            LENGTH &            N\_BAD\_Q &        Prot\_ACF\_20 \\
       Prot\_ACF\_55 &        Prot\_ACF\_80 &         Prot\_CS\_20 \\
        Prot\_CS\_55 &         Prot\_CS\_80 &       Prot\_GWPS\_20 \\
      Prot\_GWPS\_55 &       Prot\_GWPS\_80 &         START\_TIME \\
        Sph\_ACF\_20 &         Sph\_ACF\_55 &         Sph\_ACF\_80 \\
         Sph\_CS\_20 &          Sph\_CS\_55 &          Sph\_CS\_80 \\
       Sph\_GWPS\_20 &        Sph\_GWPS\_55 &        Sph\_GWPS\_80 \\
              Teff   &              kepmag &               logg \\
\hline
\end{tabular}
\end{adjustbox}

\end{table}

\section{ROOSTER performance with simulated data \label{appendix:benchmark_simulations}}

\citet{2015MNRAS.450.3211A} performed a hare and hounds exercise with simulated data. Several teams participated to the exercise, each one with their own methodology. The working sample was constituted of 1000 simulated light curves and five 1000-day solar light curves obtained with the Variability of Solar Irradiance and Gravity Oscillations \citep[VIRGO,][]{1995SoPh..162..101F} instrument on board the Solar and Heliospheric Observatory \citep[SoHO,][]{1995SoPh..162....1D}. Among the 1000 simulated light curves, noise from real \textit{Kepler} observations was added to 750, while the other 250 were kept noise-free.  

The ROOSTER training methodology presented in this paper was slightly modified to be applied on the simulated data. 
The training set was based on the same K and M stars from S19. However, five stars were removed from the training set because they were also used as noise sources for the simulated light curves in \citet{2015MNRAS.450.3211A}.
\textit{RotClass} and \textit{PeriodSel} were trained without the stellar global parameters ($T_\mathrm{eff}$ and $\log g$), and the FliPer values. Furthermore, we also abandon the Kp, bad quarter flags, number of bad quarters in the light curves, lengths, start and end date of the light curves. 
Note that ROOSTER was applied blindly to mimic the real working situation and the outputs were compared to the correct rotation periods only at the end.

Table~\ref{tab:hare_and_hounds} compares the ROOSTER results to those obtained by the CEA team in \citet{2015MNRAS.450.3211A}. Note that in \citet{2015MNRAS.450.3211A}, a previous version of our rotation pipeline (see Appendix A for details on the version used in the current study) was used. For both (noisy and noise-free data), the ROOSTER accuracy (83.2$\%$ compared to 88$\%$) is slightly lower than previously. Nevertheless, ROOSTER provided \textit{good} rotation periods \citep[i.e. periods within 10\% of the median of the observable periods $P_\mathrm{obs}$ defined in][]{2015MNRAS.450.3211A} for 73.9\% of the noisy light curves and 80.4\% of the noise-free light curves. These results represent an improvement in comparison with the results from any method used in the hare and hounds exercise. In 2015, the CEA team had the best scores, with respectively 68.6\% of good periods provided for noisy light curves and 75.4\% for noise-free light curves. 

\begin{figure}[h]

    \centering
    \includegraphics[width=0.48\textwidth]{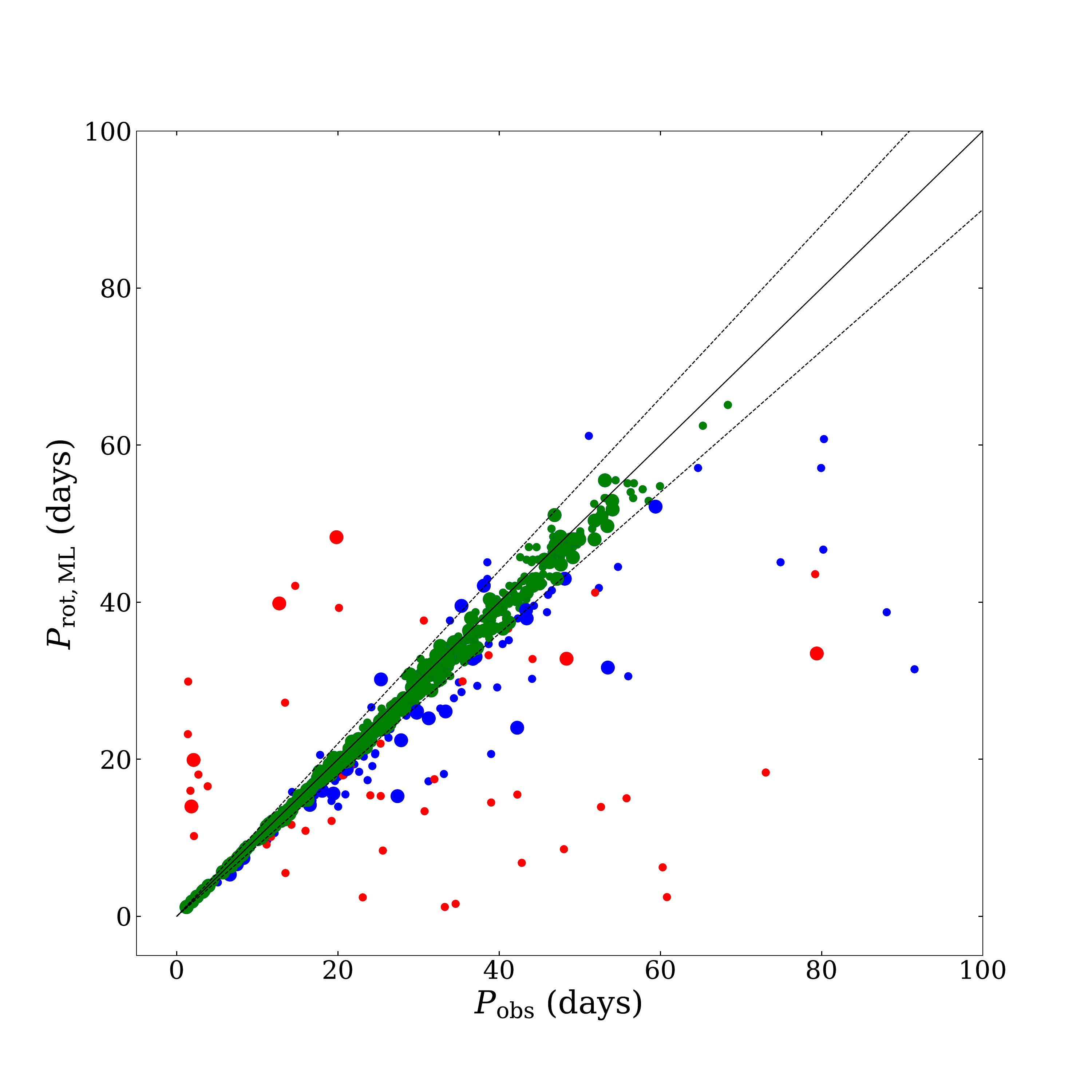}
    \caption{ROOSTER retrieved periods $P_\mathrm{rot, ML}$ versus $P_\mathrm{obs}$ from \citet{2015MNRAS.450.3211A}. \textit{Good} periods are shown in green, \textit{ok} in blue, and \textit{bad} in red. Large circles mark noise-free light curves, while small circles mark noisy light curves.}
    \label{fig:comparison_aigrain}
\end{figure}

In the hare and hounds exercise \citep{2015MNRAS.450.3211A}, stars were flagged as \textit{ok} if the detected rotation period (also considering the uncertainty) was \textit{good} or within the range of observable periods $[P_\mathrm{obs, min}, P_\mathrm{obs, max}]$. A non-\textit{ok} rotation period is denoted \textit{bad}. Following this approach, ROOSTER was able to provide \textit{ok} rotation periods for 83\% of the noisy light curves and for 90\% of the noise-free ones. Indeed, ROOSTER performs better than any method in the hare and hounds exercise. 
The comparison between ROOSTER periods $P_\mathrm{rot,ML}$ and the reference values $P_\mathrm{obs}$ is shown in Fig.~\ref{fig:comparison_aigrain}.

Concerning the solar light curves, ROOSTER detected four rotation periods among the five. Two of those rotation periods are within the 25-30 days range and are therefore considered as correct. 

As a final exercise, we decided to flag the light curves for visual check, similarly to what is described in Sect.~\ref{section:discussion}. Light curves with a mean classification ratio between 0.4 and 0.8 were flagged. Additionally, from the subsample for which ROOSTER provided a period, we flagged (for details, see in Sect.~\ref{section:discussion}):
\begin{enumerate}
    \item light curves for which $P_\mathrm{rot,ML}$ might be an harmonic of the actual rotation period;
    \item light curves with $P_\mathrm{rot,ML} > 38$ days or $P_\mathrm{rot,ML} < 1.6$ days; 
    \item light curves with a rotation period estimate not belonging to the proper filter.
\end{enumerate}

This procedure led to 362 light curves being flagged among the \citet{2015MNRAS.450.3211A} sample of 1005 light curves. 41 of the noisy light curves and 9 of the noise-free light curves without detected period would have been proposed for visual inspection. Among the light curves with detected period, 36 noisy light curves and 7 noise-free light curves with \textit{bad} period would have been visually checked. Note that the visual check does not guarantee that the ROOSTER wrong determinations would be corrected.  


\begin{table*}[h]
    \centering
    \caption{Comparison between the 2015 CEA team results in the hare and hound exercise and the ROOSTER results from the simulated data. For each sample (noisy, noise-free and solar), percentage of detected (det) rotation periods is given, followed by percentage of \textit{good} rotation periods among the detected rotation period and the full sample. The same percentages are also provided for \textit{ok} rotation periods. The \textit{good} and \textit{ok} nomenclature in described in detail in \citet{2015MNRAS.450.3211A} as well as in Appendix~\ref{appendix:benchmark_simulations}.}
    \label{tab:hare_and_hounds}
    \begin{tabular}{l|rrrrr|rrrrr|rr}
       \hline\hline
       Method & \multicolumn{5}{c}{Noisy} &
       \multicolumn{5}{c}{Free} & \multicolumn{2}{c}{Solar} \\
       
       & \% det & \multicolumn{2}{c}{\% good} & \multicolumn{2}{c}{\% ok} & \% det & \multicolumn{2}{c}{\% good} & \multicolumn{2}{c}{\% ok} & No. det & No. ok \\
       \hline
       & & det & global & det & global & & det & global & det & global & &  \\
       CEA 2015   & 78 & 88 & 68.6 & 95 & 74.1 &  82 & 92 & 75.4 & 99 &  81.2 & 2 & 2 \\
       ROOSTER   & 88.8 & 83.2 & 73.9 & 93.5 & 83 & 92.8 & 86.6 & 80.4 & 97 & 90 & 4 & 2 \\
       \hline
    \end{tabular}

\end{table*}

\section{Output format and visual check-flags signification}

The standard ROOSTER output files are saved as comma-separated values (csv) files. The column order of the standard ROOSTER output files is given in Table~\ref{tab:columns_order}, while the signification of the visual checks flag is summarised in Table~\ref{tab:flag_signification}. The flags follow a hierarchical order, i.e., if there is an overlap of flags, the table will provide solely the first flag to be considered.. 

\begin{table}[h]
    \caption{Column order of the output ROOSTER files.}
    \label{tab:columns_order}
\centering
\begin{tabular}{lc}
\hline\hline
KIC                      &     - \\
label \textit{RotClass}  &     - \\
$P_\mathrm{rot, ML}$      &  days \\
$P_\mathrm{rot,error}$    &  days \\
\sph                     &     - \\
$S_\mathrm{ph, error}$   &     - \\
flag for visual check    &     (see Table~\ref{tab:flag_signification}) \\
label \textit{FlagPoll}  &     - \\
classification ratio \textit{Rot}    &     - \\
classification ratio \textit{FlagPoll} &     - \\
flag missing parameters  &     - \\
$T_\mathrm{eff}$          &     K \\
$\log g$                 &   dex \\
\hline
\end{tabular}

\end{table}

\begin{table}[h]
\caption{Meaning of visual-check flags in ROOSTER output files.}
\label{tab:flag_signification}
\centering
\begin{tabular}{c|c}
\hline\hline
-1 & $P_\mathrm{rot}$ corrected by re-attributing filter (no check needed) \\
0  & no check needed \\
10 & harmonic candidate \\
12 & instrumental modulation candidate ($P_\mathrm{rot}$ > 38 days) \\
14 & filter \\
16 & 0.4 < classification ratio \textit{Rot} < 0.8 \\
18 & Type 1 CP/CB candidate with $P_\mathrm{rot}$ > 7 days \\
20 & $P_\mathrm{rot}$ < 1.6 days \\
22 & observational length shorter than quarters \\
\hline
\end{tabular}

\end{table}

\end{document}